\documentclass[12pt]{article}
\usepackage{amsfonts,amssymb}

\def\bq{ \begin{equation} }
\def\eq{ \end{equation} }
\def\ben{ \begin{eqnarray} }
\def\en{ \end{eqnarray} }

\def\frac#1#2{{#1\over #2}}

\def\on#1#2{\mathop{\vbox{\ialign{##\crcr\noalign{\kern2pt}
$\scriptstyle{#2}$\crcr\noalign{\kern2pt\nointerlineskip}
\kern-2pt$\hfil\displaystyle{#1}\hfil$\crcr}}}\limits}

\begin{document}

\baselineskip=15pt
\vspace{1cm} \noindent {\LARGE \textbf {On
 (2+1)-dimensional hydrodynamic type \\[3mm] systems possessing  pseudopotential  with \\[3mm] movable singularities}}
\vskip1cm \hfill
\begin{minipage}{13.5cm}
\baselineskip=15pt {\bf
Alexander Odesskii ${}^{1,} {}^{2}$ and
Vladimir Sokolov ${}^{1}$} \\ [2ex]
{\footnotesize ${}^1$ Landau Institute for Theoretical Physics,
Moscow, Russia
\\
${}^{2}$  School of Mathematics, The University of
Manchester, UK}\\
\vskip1cm{\bf Abstract}

A certain class of integrable hydrodynamic type systems with three
independent and $N\ge 2$ dependent variables is considered. We
choose the existence of a pseudopotential as a criterion of
integrability. It turns out that the class of integrable systems
having pseudopotentials with movable singularities is described by a
functional equation, which can be solved explicitly.  This  allows
us to construct interesting examples of integrable hydrodynamic
systems for arbitrary $N$.

\end{minipage}

\vglue1cm \textbf{Address}: Landau Institute for Theoretical
Physics, Kosygina 2, 119334, Moscow, Russia

\textbf{E-mail}: alexander.odesskii@manchester.ac.uk, \,
sokolov@itp.ac.ru
\newpage

\centerline{\Large\bf Introduction}
\medskip

In the papers \cite{ferhus1, ferhus2, ferhus3} a general theory of integrable systems of PDEs of the form
 \begin{equation}   \label{fergener}
 {\bf u}_t=A({\bf u})\, {\bf u}_x+ B({\bf u}) \, {\bf u}_y,
  \end{equation}
where ${\bf u}$ is an $N$-component column vector,  $A({\bf u})$ and
$B({\bf u})$ are $N\times N$-matrices, was developed. The existence
of sufficiently many of the hydrodynamic reductions \cite{gibtsar,
ferhus1} has been proposed as the definition of integrability.
Unfortunately, for arbitrary $N$ it is difficult to write down
explicitly the conditions for $A$ and $B,$ which follows from this
definition. Nevertheless, for $N=2$ the complete set of
integrability conditions have been found in the paper
\cite{ferhus2}.  For $N>2$ even the verification whether a given
equation is integrable is a serious task. Any attempt for classification of
integrable models based directly on this definition seems to be
hopeless.

To overcome the difficulties for $N>2$ the following two
observations \cite{ferhus2, ferhus3} are very useful. First, under
some conditions of generic position, the matrix $M=(1+t A)^{-1}(1+k
B)$ for integrable models should be diagonalizable by a point
transformation ${\bf u}\rightarrow \Phi({\bf u})$ for  generic
values of the parameters $t$ and $k$. If the eigen-values of $M$ are distinct, this
is equivalent to the fact that the Haantjes tensor \cite{hant} of
$M$ is identically zero. This gives rise to an overdetermined system
of the first order PDEs for entries of $A$ and $B$. Given a system
(\ref{fergener}) it is not difficult to verify  whether
these PDEs are satisfied or not. The simplest equations from this
overdetermined system also can be useful for classification of
integrable models (\ref{fergener}).

The second observation made in \cite{ferhus1} is that for $N=2$ the integrability conditions are equivalent
to the existence of the scalar pseudopotential
\begin{equation}\label{pseudo0}
 \Psi_t=f(\Psi_y, {\bf u}), \qquad   \Psi_x=g(\Psi_y, {\bf u}),
\end{equation}
for (\ref{fergener})\footnote{The latter means that that the
overdetermined system (\ref{pseudo0}) for $\Psi$ is compatible if
and only if ${\bf u}$ is a solution of (\ref{fergener}). }. The
scalar pseudopotential plays an important role in the theory of the
universal Whitham hierarchy \cite{kr1,kr3,kr4}. A possible
importance of pseudopotentials was also noticed in \cite{zakh}. The
existence of pseudopotential implies a representation of
(\ref{fergener}) as the commutativity condition for the
corresponding characteristic vector fields
$$
 \frac{\partial}{\partial x}-g_{\xi} \frac{\partial}{\partial y}+g_{y} \frac{\partial}{\partial \xi}, \qquad
  \frac{\partial}{\partial t}-f_{\xi} \frac{\partial}{\partial y}+f_{y} \frac{\partial}{\partial
  \xi}, \qquad \mbox{where} \qquad \xi=\Psi_y.
$$
For recent attempts to use
similar representations for integration of dispersionless PDEs see
\cite{mansan1,mansan2}.

In this paper we assume that the matrix $A({\bf u})$ is constant and
consider integrable systems of the form
\begin{equation}   \label{gener}
 u_{it}=\lambda_i \, u_{ix}+ \sum_{1\leq j\leq N} b_{ij}({\bf u})\,u_{j y},
\end{equation}
where $\lambda_1,\dots,\lambda_N $ are pairwise distinct constants. Functions $b_{ij}$
(as well as all other functions) are supposed to be locally analytic.
Note that the transformation
\begin{equation}   \label{subs}
u_i\to \psi_i(u_i)
\end{equation}
preserves the form of the system (\ref{gener}) for arbitrary
functions of one variable $\psi_i(u_i)$.

For $N=2$ such systems
were considered in \cite{ferhus2}. Our goal is to obtain a list of the most interesting examples of integrable
models (\ref{gener}) with $N>2$.
As far as we know, nobody systematically investigated such systems before us.

{\bf Example 1.} Consider equation  (\ref{gener}) with
$$
 b_{ij}=\frac{\lambda_i-\lambda_j}{u_i-u_j} \, c_j, \qquad i\ne j,
$$
\begin{equation}   \label{bii}
 b_{ii}=-\sum_{j\ne i} b_{ij},
\end{equation}
where $c_j$ are arbitrary constants.
It is not difficult to verify that for arbitrary $N$ this equation
possesses pseudopotential (\ref{pseudo0}), where
\begin{equation}   \label{ex1}
g=\sum_{i=1}^N  c_i  \log (u_i- \Psi_y), \qquad  f=\sum_{i=1}^N  c_i \lambda_i  \log (u_i- \Psi_y).
\end{equation}
This pseudopotential has the following structure:
\begin{equation}   \label{potensum}
g=  \sum_{i=1}^N h_i(\xi, u_i), \qquad  f=  \sum_{i=1}^N \lambda_i h_i(\xi, u_i),
\end{equation}
where $\xi=\Psi_y$.
In Section 2 we show that (\ref{potensum}) is true for any integrable equation of the form (\ref{gener}).

Notice that functions $h_i(\xi, u_i)$ in (\ref{ex1}) have "moveable" singularities with respect to the variable
$\xi=\Psi_y$. This means that the position of the singularity depends on ${\bf u}.$

In this paper we describe all equations (\ref{gener}) possessing pseudopotentials such that for any $i$ the function
$h_i(\xi, u_i)$ has a movable singularity.  This leads to series of new interesting examples of integrable systems
(\ref{gener}).  These examples are presented in Section 1.
It would be interesting to find the hydrodynamic reductions for these equations and describe
the multiple waves \cite{perad, ferhus1} in terms of these reductions.

In Sections 2-4 we deduce a functional equation describing the pseudopotentials with moveable singularities
and find all it's solutions. These solutions correspond to examples of Section 1 and their
degenerations.

\section{Examples.}

{\bf Example 2.} Consider equation  (\ref{gener}) given by
$$
 b_{ij}=c_j (\lambda_j-\lambda_i) \left(\kappa+\frac{e^{u_i-u_j} }{e^{u_i-u_j}-1 }\right),       \qquad i\ne j,
$$
and
$$
 b_{ii}=-\sum_{j\ne i} b_{ij}.
$$
Here if $\kappa=-1$ or $\kappa=0,$ then $c_j$ are arbitrary constants. For
other $\kappa$ the constants $c_j$ should satisfy the following two
relations
\begin{equation}   \label{const}
\sum_{i=1}^N c_i=0, \qquad \sum_{i=1}^N \lambda_i c_i=0.
\end{equation}
It is easy to verify that for any $N$ this equation admits a pseudopotential (\ref{potensum}) with
\begin{equation}   \label{ex2}
h_i(\xi, u)=c_i \Big(\kappa (\xi-u)+\log(e^{\xi-u}-1) \Big).
\end{equation}

{\bf Example 3.} Consider equation  (\ref{gener}) with
$$
 b_{ij}=c_j (\lambda_j-\lambda_i) \frac{u_i-u_j+1}{u_i-u_j}\,\cdot \, \frac{u_i}{u_j},       \qquad i\ne j,
$$
$$
 b_{ii}=\sum_{j\ne i} c_j (\lambda_j-\lambda_i) \left(\frac{1}{u_j-u_i}
 +\log(u_j)\right),
$$
where $c_j$ are arbitrary constants satisfying conditions
(\ref{const}).
This equation has a pseudopotential (\ref{potensum}) with
\begin{equation}   \label{ex3}
h_i(\xi, u)=c_i \Big((\xi+1) \log(u)-\log(u-\xi) \Big).
\end{equation}

The equation from Example 1 is a particular case of the following
model:

{\bf Example 4.} Let
$$
 b_{ij}=\frac{(\lambda_i-\lambda_j) c_j  P(u_i)  M(u_j)}{u_i-u_j} , \qquad i\ne j,
$$
$$
 b_{ii}=-\sum_{j\ne i} \frac{(\lambda_i-\lambda_j) c_j
 P(u_j) M(u_j)}{u_i-u_j}-\sum_{j\ne i} (\lambda_i-\lambda_j) c_j B(u_j),
$$
where $c_j$ are arbitrary constants, and  the functions $B, M$ are defined by quadratures from
$$
 B'= (k_3 x+k_2 + z_1)\,M, \qquad
 M'=M \frac{-k_3 x^2+z_1 x+z_0}{P}.
$$
Here
$$
 P(x)=k_3 x^3+k_2 x^2+k_1 x+k_0
$$
is an arbitrary polynomial of degree not greater than 3, $z_1,z_0$
are arbitrary constants.

The corresponding equation (\ref{gener})  possesses
pseudopotential (\ref{potensum}), where
$$
h_u(\xi,u)= -\frac{M(u)}{u-\phi(\xi)}\cdot\frac{1}{M(\phi(\xi))},
\qquad h_{\xi}(\xi,u)=-B(u)+\frac{P(u) M(u)}{u-\phi(\xi)},
$$
 $$
 \phi'= P(\phi)\, M(\phi) .
 $$
 For any given $P$ and $z_1, z_0$ the equations for $\phi(\xi)$ and $h(\xi,u)$ can be
 easily solved by quadratures.

Using admissible transformations
$$
u_i \rightarrow \frac{a u_i+b}{c u_i+d},   \qquad  i=1,\dots,N,
$$
one can reduce the polynomial P to a canonical form. For example, if
all three roots of $P$ are distinct, then without loss of generality we
may put $P(x)=x(x-1).$ In this case
$$
 M(x)= x^{s_1} (x-1)^{s_2} ,
$$
where $s_1=-z_0, s_2=z_0+z_1,$ and
$$
 B(x)=(s_1+s_2+1) \int x^{s_1} (x-1)^{s_2} dx .
$$
It is not difficult to find that
$$h(\xi,u)=\frac{1}{\phi(\xi)^{s_1}(\phi(\xi)-1)^{s_2}}\int_c^u\frac{t^{s_1}(t-1)^{s_2}}{\phi(\xi)-t}dt
-c^{s_1+1}(c-1)^{s_2+1}\int\frac{d\xi}{\phi(\xi)-c},$$ where
$\phi'=\phi^{s_1+1}(\phi-1)^{s_2+1}$.

Other two canonical forms are $P=x$ and $P=1$. The latter generates
Example 1 if $z_1=z_0=0.$

\section{Pseudopotentials.}

A pair of equations of the form
\begin{equation}\label{pseudo}
 \Psi_t=f(\Psi_y, u_1,\dots, u_N), \qquad   \Psi_x=g(\Psi_y, u_1,\dots, u_N),
\end{equation}
with respect to unknown $\Psi$ is called a pseudopotential for
equation (\ref{gener}) if the compatibility  condition
$\Psi_{tx}=\Psi_{xt}$ for (\ref{pseudo}) is equivalent to
(\ref{gener}). Differentiating (\ref{pseudo}), we find that this
compatibility condition is given by
$$
f_{\xi}\, \sum_{i=1}^N (u_i)_y \,\partial_i g   + \sum_{i=1}^N  (u_i)_x\,\partial_i f =
g_{\xi} \sum_{i=1}^N (u_i)_y \, \partial_i f   + \sum_{i=1}^N (u_i)_t \,\partial_i g.
$$
Here and below we denote $\Psi_y$ by $\xi$ and
$\frac{\partial}{\partial u_i}$ by $\partial_i$. Substituting the
right hand side of (\ref{gener}) for $t$-derivatives and splitting
with respect to $x$ and $y$-derivatives, we get that for any $i$ the following relations hold:
\begin{equation}\label{psecon1}
 \partial_i f=\lambda_i \partial_i g,
 \end{equation}
 \begin{equation}\label{psecon2}     f_{\xi} \, \partial_i
 g - g_{\xi} \, \partial_i f=\sum_{j=1}^N b_{ji} \partial_j g.
\end{equation}
Since $\lambda_i$ are pairwise distinct, it follows from the
condition (\ref{psecon1}) that
$$
g=  \sum_{i=1}^N h_i(\xi, u_i),
$$
and
$$f=  \sum_{i=1}^N \lambda_i h_i(\xi, u_i)+c(\xi).
$$
It is easy to see that the integration constant $c(\xi)$ can be distributed between
functions $h_i.$ Thus we have arrived at  (\ref{potensum}).

Substituting (\ref{potensum}) into (\ref{psecon2}), we obtain
\begin{equation}   \label{main1}
\partial_i h_i(\xi,u_i)  \sum_{j} (\lambda_j- \lambda_i)
h_{j\,\xi}(\xi,u_j) = \sum_{j} b_{ji}\,\, \partial_j h_j(\xi,u_j) .
\end{equation}

{\bf Remark 1.} If we fix a generic value $\xi_0$ of the variable
$\xi$, then it follows from (\ref{main1}) that
   \begin{equation}   \label{biii}
 b_{ii}=  \sum_{j} (\lambda_j-\lambda_i) \phi_j(u_j) - \sum_{j\ne i} b_{ji}
 \frac{s_j(u_j)}{s_i(u_i)},
    \end{equation}
where $\phi_j(u_j)=h_{j\,\xi}(\xi_0,u_j),$ $s_j(u_j)=\partial_j
h_j(\xi_0,u_j).$

\section{Basic functional equation.}

Suppose that for any $i$ the function $h_i(\xi,u)$ has a singularity
with respect to $\xi$ and this singularity depends on $u$. After  a
transformation of the form (\ref{subs}), we may assume that for each $i$ the
singularity of $h_i(\xi,u)$ is located on the diagonal $\xi=u$.
We say that $h_i(\xi,u)$ has a singularity on the diagonal, if for fixed generic $u$ and
any $\epsilon>0$ we have $\max\{|h_i(\xi,u)|,
|\xi-u|<\epsilon\}=\infty$.

{\bf Proposition 1.} {\it Suppose} $h_i(\xi,u)$ {\it has a singularity on the diagonal}
$\xi=u$ {\it for each} $i$. {\it Then there exist: a function} $h(\xi,u)$,
{\it functions}  $f_i(\xi)$ {\it and non-zero constants} $c_i$
{\it such that}
\begin{equation}   \label{1}
h_i(\xi,u)=c_ih(\xi,u)+f_i(\xi),
\end{equation}
\begin{equation}   \label{2}
b_{ji}=(\lambda_i-\lambda_j) c_i \, \partial_i h(u_j,u_i), \qquad
i\ne j,
\end{equation}
\begin{equation}   \label{3}
b_{ii}=\sum_{j\ne i}(\lambda_j-\lambda_i)(c_j\, \partial_i
h(u_i,u_j)+f^{\prime}_j(u_i)),
\end{equation}
\begin{equation}   \label{4}
h(\xi,u)=\ln(\xi-u)+{\rm regular \, part}.
\end{equation}
{\it Moreover, the following functional equation}
\begin{equation}   \label{funceq}
h_{\xi}(\xi,w)\,
h_v(\xi,v)+h_v(w,v)\, h_w(\xi,w)-h_v(v,w)\, h_v(\xi,v)=\nu(\xi,v)
\end{equation}
{\it holds  for some function}  $\nu$.

{\bf Proof}. Considering (\ref{main1}) near the diagonal $\xi=u_j$, where
$j\ne i,$ and comparing the singularities, we obtain
\begin{equation}   \label{bij}
b_{ji}=(\lambda_j-\lambda_i)\mu_j(u_j)\, \partial_i h_i(u_j,u_i),
\end{equation}
where $\mu_j(u_j)=\lim_{\xi\rightarrow
u_j}\frac{h_{j\,\xi}(\xi,u_j)}{\partial_jh_j(\xi,u_j)}$. We see that
the function $b_{ji}$ depends on $u_i,u_j$ only  and has the same
singularity on the diagonal as $\partial_i h_i(u_j,u_i)$.
Considering (\ref{main1}) near the diagonal $u_i=u_j$, comparing the
singularities and using (\ref{biii}), we obtain $\partial_jh_j(\xi,
u_i) s_i(u_i)=\partial_ih_i(\xi, u_i) s_j(u_i)$ or
$\partial_ih_i(\xi, u_i)=\nu_i(u_i)r(\xi,u_i)$ for some functions
$\nu_i$ and $r$. On the other hand, consider (\ref{main1}) for $N$
generic values $\xi_1,...,\xi_N$ of variable $\xi$.  For each fixed
$i=1,...,N$ we have a system of $N$ linear equations for
$b_{i1},...,b_{iN}$ with matrix $Q=(q_{jk}),$ where
$q_{jk}=\nu_j(u_j)r(\xi_k,u_j) .$ This system must have a unique
solution by definition of pseudopotential and therefore $\Delta=\det
Q\ne0$. It is clear that $b_{ji}=\frac{P_{ji}}{\Delta},$ where
$P_{ji}$ is regular on each diagonal $u_k=u_l.$ It is easy to prove
the following

{\bf Lemma.} Let $\Delta(u_1,...,u_m)$ be the determinant of an $m\times m$ matrix $Q$,
whose entries $q_{ij}$ have the form  $q_{ij}=g_i(u_j)$ for
some functions $g_1,...,g_m$. The function $\Delta$ is not equal to
zero identically iff the functions $g_1,...,g_m$ are linearly  independent.
In this case $\partial_i\Delta\ne 0$ on the diagonal
$u_i=u_j$ for each $i\ne j$.

From this lemma it follows that the only singularity of
$b_{ij}$ on the diagonal can be a pole of order one. Taking into
account (\ref{bij}), we obtain that near $u_j=u_i$
$$\partial_i h_i(u_j,u_i)=\frac{\alpha_i(u_j)}{u_j-u_i}+{\rm regular \,\, part}
$$
or, after integration,
$$h_i(u_j,u_i)=-\alpha_i(u_j)\ln(u_j-u_i)+{\rm regular \,\, part}.$$
Considering the singular part of (\ref{main1}) at   $\xi=u_j$,
we obtain $\mu_j(u_j)=-1$ and $\alpha_i^{\prime}(u_j)=0,$ i.e.
$\alpha_i(u_j)=-c_i$ for some constant $c_i$. Comparing the singularities in (\ref{main1})
at $\xi=u_i$,  we find that
$$b_{ii}=\sum_{j\ne
i}(\lambda_j-\lambda_i)\, \partial_i h_j(u_i,u_j).$$ Substituting this
expression for $b_{ii}$ into (\ref{main1}), we obtain
\begin{equation}   \label{main2}
\sum_{j\ne
i}(\lambda_j-\lambda_i)\,\Big( h_{j\,\xi}(\xi,u_j)\,  \partial_i h_i(\xi,u_i)
+\partial_i h_i(u_j,u_i) \, \partial_j h_j(\xi,u_j) -
\partial_i  h_j(u_i,u_j) \, \partial_i h_i(\xi,u_i)\Big) =0.
\end{equation}
Considering the singular part of (\ref{main2}) at $u_i=u_j,$ we get
$$c_j(h_i(\xi,u))_u=c_i(h_j(\xi,u))_u,$$ which gives (\ref{1}) and (\ref{4}).
Now (\ref{2}) and (\ref{3}) follow from (\ref{1}) and expressions for $b_{ij}$ and $b_{ii}$  already obtained.
  Using (\ref{main2}), we arrive at the relation
$$ \begin{array}{c}
\sum_{j\ne
i}(\lambda_j-\lambda_i)c_j\, \Big(h_{\xi}(\xi,u_j)\, \partial_i h(\xi,u_i)
+\partial_i h(u_j,u_i) \partial_j h(\xi,u_j) -\partial_i h(u_i,u_j)\partial_i h(\xi,u_i)
+\\[4mm]
(f_j^{\prime}(\xi)-f_j^{\prime}(u_i))\,\partial_i h(\xi,u_i)\Big)=0.
\end{array}$$
This relation gives the equation (\ref{funceq}) for some function
$\nu(\xi,v)$. $\blacksquare$

{\bf Remark 2.} If a pair $h(\xi,v)$, $\nu(\xi,v)$ is a solution of
(\ref{funceq}), then
\begin{equation}   \label{subst}
\tilde{h}(\xi,v)=h(\xi,v)+f(\xi) ,  \qquad
\tilde{\nu}(\xi,v)=\nu(\xi,v)+(f^{\prime}(\xi)-f^{\prime}(v))h_v(\xi,v)
\end{equation}
is also a solution of (\ref{funceq}). Therefore, if $\nu(\xi,v)$ has
a form $(g(\xi)-g(v))h_v(\xi,v)$ for some function $g$, then we can
bring $\nu(\xi,v)$ to $0$ adding to $h(\xi,v)$ a suitable function of
$\xi$.

{\bf Remark 3.} Without loss of generality, we can assume that $f_i(\xi)=f_j(\xi)$ for all $i,j$.
 Indeed, only the linear combinations
$\sum_if_i(\xi)$ and $\sum_i\lambda_if_i(\xi)$ appear  in (\ref{potensum}). Furthermore, according to Remark 2,
we may put $f_i(\xi)=0$.

{\bf Remark 4.} It follows from (\ref{2}), (\ref{funceq}) that for any $i\ne j$  the
function $b=b_{ij}(u_i,u_j)$ satisfies the following functional
equation
$$
 b(w,v) b_w(x,w)-b(x,v) b_v(v,w)+b(x,w) b_w(w,v)+b(x,v)
 b_x(x,w)=0.
$$

{\bf Remark 5.} It follows from (\ref{2}), (\ref{3}) that the equation (\ref{gener})
defined by (\ref{2}) and  (\ref{3}) with $f_i(\xi)=0$
can be written in the following divergent form:
$$
 u_{it}=\lambda_i \, u_{ix}+ \sigma_{iy},
$$
where
$$
\sigma_i=\sum_{j\ne i} (\lambda_j-\lambda_i) c_j h(u_i,u_j).
$$
Thus any such equation has at least $N$ linearly independent hydrodynamic conservation laws.

{\bf Proposition 2.} {\it Let} $h(\xi,v)$ {\it be a solution of} (\ref{funceq})
{\it with} $\nu(\xi,v)=0$, {\it then for any nonzero constants} $c_i$ {\it the formula}
\begin{equation}   \label{pspot}
g=  \sum_{i=1}^N c_ih(\xi, u_i),\qquad f=  \sum_{i=1}^N \lambda_i
c_ih(\xi, u_i)
\end{equation}
{\it defines a pseudopotential for equation} (\ref{gener}) {\it given by}
(\ref{2}) {\it and} (\ref{3}) {\it with} $f_i(\xi)=0$.

{\it Let} $h(\xi,v)$ {\it be a solution of} (\ref{funceq}) {\it
with} $\nu(\xi,v)\ne 0$, {\it then} (\ref{pspot}) {\it defines
pseudopotential for equation} (\ref{gener}) {\it given by} (\ref{2}) {\it and} (\ref{3})
{\it with} $f_i(\xi)=0$ {\it iff
 the constants} $c_i$ {\it satisfy the relations} (\ref{const}).

{\bf Proof.} According to the previous results, (\ref{pspot})
defines a pseudopotential iff
$$
\sum_{j\ne
i}(\lambda_j-\lambda_i)c_j\Big( h_{\xi}(\xi,u_j)\,\partial_i h(\xi,u_i)
+\partial_i h(u_j,u_i)\,\partial_j h(\xi,u_j)-\partial_i h(u_i,u_j)\, \partial_i h(\xi,u_i) \Big)=0.
$$
Substituting (\ref{funceq}) into this relation, we obtain the statement of the
proposition. $\blacksquare$

\section{Classification of solutions for the functional equation.}

{\bf Proposition 3.} {\it Let a pair} $h(\xi,v)$, $\nu(\xi,v)$ {\it be a
solution of} (\ref{funceq}) {\it with asymptotic} (\ref{4}). {\it Then up to
substitutions of the form} (\ref{subst}) {\it it belongs to the following list}:
$$
 h(x,v)=\kappa \,(x-v)+\log( x-v ), \qquad \nu(x,v)=\kappa (\kappa+1) ;
$$
$$
 h(x,v)=\kappa \,(x-v)+\log(e^{x-v}-1), \qquad \nu(x,v)=\kappa (\kappa+1),
$$
{\it where} $\kappa$ {\it is an arbitrary constant};
$$
h(x,v)=(x+1) \log(v)-\log(v-x), \qquad  \nu(x,v)=\frac{x}{v}  ;
$$
$$
h(x,v)=\int_c^v\frac{P(\phi(x))\,\phi^{\prime}(t)^2}{(\phi(x)-\phi(t)) \,
P(\phi(t))\,\phi^{\prime}(x)}dt-\int\frac{\phi^{\prime}(c)}{\phi(x)-\phi(c)}dx
, \qquad  \nu(x,v)=0.
$$
{\it Here} $c$ {\it is a constant and the function}  $\phi$ {\it is defined by the
following differential equation}:
$$\phi''= \Big(\frac{2 P'(\phi)}{3 P(\phi)}+ \frac{Z(\phi)}{P(\phi)}
\Big)  \,\phi'\,^2  ,
$$
{\it where}
$$
 P(x)=k_3 x^3+k_2 x^2+k_1 x+k_0,  \qquad Z(x)=z_1 x+z_0
$$
{\it are arbitrary polynomials such that} {\rm \, deg} $P\le 3$ {\it and} {\rm  \, deg}
$Z\le 1.$

{\bf Proof.} According to (\ref{4}), we have an expansion of the
form
\begin{equation}   \label{asimp}
h(w,v)=\ln(w-v)+\sum_{i=0}^{\infty} a_i(w) (w-v)^i
\end{equation}
as $v$ tends to $w.$
To describe the solutions of the functional equation (\ref{funceq}), let us
investigate a set of conditions   relating the functions $a_i.$ Using expansion (\ref{asimp}) for $h(v,w)$ and
$h(w,v)$ and equating the coefficients of different powers of $v-w,$
we obtain an infinite sequence of PDEs for the function $h(x,v)$ and the coefficients $a_i(v)$.
The simplest three of these PDEs read as follows:
\begin{equation}   \label{Eq1}
 h_{vvv}-2 h_v h_{x v}+2 a_1  h_{vv}=0,
\end{equation}
\begin{equation}   \label{Eq2}
 h_{vvvv}-3 h_v h_{xvv}+3 a_1  h_{vvv}+6 (a_1'+2 a_2) h_{vv}+3 (a_1''+6 a_2'+12 a_3)
 h_v=0,
\end{equation}
\begin{equation}  \begin{array}{c} \label{Eq3}
 h_{vvvvv}-4 h_v h_{xvvv}+4 a_1  h_{vvvv}+12 (a_1'+2 a_2) h_{vvv}+
 12 (a_1''+4 a_2'+6 a_3) h_{vv}\\[3mm]+4 (a_1'''+6 a_2''+12 a_3')
 h_v=0.
\end{array}
\end{equation}
Substituting expansion (\ref{asimp}) for $h(x,v)$ to
(\ref{Eq1}), we observe that all coefficients $a_i, \,\,i>2$ can be
expressed as certain differential polynomials of $a_1$ and $a_2$.
For example,
$$a_3=-\frac{1}{12} (a_1''+2 a_1 a_1'+4 a_2').
$$
This means that the function $h(x,v)$ is uniquely determined by functions
$a_1(v)$ and $a_2(v)$.

The expansion of equations (\ref{Eq2}) and (\ref{Eq3}) leads to
differential relations between $a_1$ and $a_2$. In particular, the simplest relation
following from (\ref{Eq2}) has the form
$$
a_1'''+6 a_1 a_1''+6 a_1'\,^2-6 a_1^2 a_1' +12 (a_2 a_1'+a_2'
a_1)=0.
$$
If $a_1\ne 0,$ this implies
\begin{equation}   \label{a2}
a_2=\frac{C-a_1''-6 a_1 a_1'+2 a_1^3}{12 a_1}
\end{equation}
for some constant $C$. Eliminating $a_2$ with the help of (\ref{a2}),
we arrive at an overdetermined system of ODEs for function $a_1.$
If $C=0,$ we find from this system that
\begin{equation}   \label{dif4}
a_1^2 \,a_1^{(4)}+2 a_1 (3 a_1^2-a_1')\,a_1^{(3)}  -4 a_1\,
(a_1'')^2   +2 (a_1'\,^2-9 a_1^2 a_1'+4 a_1^4)\, a_1'' -16 a_1^3
a_1'\,^2=0.
\end{equation}
The general solution of this equation can be written as
$$
a_1(x)=-\frac{3 Z(\phi(x))}{2 P(\phi(x))}\,\phi'(x), \qquad
\phi''= \Big(\frac{2 P'(\phi)}{3 P(\phi)}+ \frac{Z(\phi)}{P(\phi)}
\Big)  \,\phi'\,^2  ,
$$
where
$$
P(x)=k_3 x^3+k_2 x^2+k_1 x+k_0,  \qquad Z(x)=z_1 x+z_0
$$
are arbitrary polynomials such that {\rm deg} $P\le 3$ and {\rm deg}
$Z\le 1.$  For any given $P$ and $Z$ the equation for $\phi$ can be
easily integrated by quadratures.   Example 4 from Section 1 after
transformation $u\rightarrow \phi(u)$ describes the pseudopotential generated
by such a function $a_1.$

Let $C\ne 0.$ Then a simple analysis of the ODE system for $a_1$ shows
that either $a_1'=0$ or
\begin{equation}   \label{difcub}
4 (a_1')^3 +12 a_1^2\, (a_1')^2+12 (a_1^4-C a_1)\,a_1'+4 a_1^6+4 C
a_1^3+C^2=0.
\end{equation}
It is easy to verify that if $a_1'=0,$ then $a_i'=0$ for any $i$ and
therefore $h(x,v)=H(x-v)$ for some function $H$. Solving equation
(\ref{Eq1}), we get
$$
  H(x)=c_1+c_2 x+\log(1-e^{-c_3 x})
$$
or
$$
H(x)=c_1+c_2 x+\log(c_3 x).
$$
These solutions correspond to the model of Example 2 and it's degeneration.

The left hand side of (\ref{difcub}) can be decomposed into three
factors. Each factor gives rise to a differential equation of the
form
\begin{equation}   \label{difkv}
a_1'+(a_1+k)^2=0,
\end{equation}
where $k$ is related to the constant $C$ from ({\ref{a2}}) by
 $C=2 k^3.$  The corresponding model is described in Example 3.

The case $a_1=0$ should be considered separately. It is easy to get
that in this case
\begin{equation}   \label{aaa2}
a_2''+36 \,a_2^2=0.
\end{equation}
It turns out that it is a particular case of the model described by (\ref{dif4}). In the
corresponding formulas one has to put $Z=0.$
$\blacksquare$

\vskip.3cm \noindent {\bf Acknowledgments.} The authors are grateful
to E.V. Ferapontov  who involved them into activity related to
classification of multi-dimensional dispersionless integrable
models, pointed out the basic role of hydrodynamic systems in the
variety of such models and described a circle of interesting open
problems on the subject. We are also grateful to M.V. Pavlov and
O.I. Mokhov for useful discussions. The second author (V.S.) is
grateful to the Manchester Institute for Mathematical Sciences for
hospitality and support. The research was also partially supported
by: EPSRC grant EP/D036178/1, RFBR grant 05-01-00189, NSh grants
1716.2003.1 and 2044.2003.2.

\end{document}